\documentclass[%
 reprint,
superscriptaddress,
 amsmath,amssymb,
 aps, prl,abbrv,
floatfix,
]{revtex4-1}

\usepackage{graphicx}
\usepackage{dcolumn}
\usepackage{bm}
\usepackage{hyperref}
\usepackage[mathlines]{lineno}

\renewcommand{\vec}[1]{\mathbf{#1}}

\begin{document}

\preprint{APS/123-QED}

\title{Theory of Linear Spin Wave Emission from a Bloch Domain Wall}

\author{N. J. Whitehead}
\affiliation{Department of Physics \& Astronomy, University of Exeter, Stocker Road, Exeter, UK, EX4 4QL.}
\author{S. A. R. Horsley}
\affiliation{Department of Physics \& Astronomy, University of Exeter, Stocker Road, Exeter, UK, EX4 4QL.}
\author{T. G. Philbin}
\affiliation{Department of Physics \& Astronomy, University of Exeter, Stocker Road, Exeter, UK, EX4 4QL.}
\author{A. N. Kuchko}
\affiliation{Institute of Magnetism, 36b Vernadskogo Avenue, Kiev, 03142, Ukraine.}
\author{V. V. Kruglyak}
\email{V.V.Kruglyak@exeter.ac.uk}
\affiliation{Department of Physics \& Astronomy, University of Exeter, Stocker Road, Exeter, UK, EX4 4QL.}

\date{\today}

\begin{abstract}
We report an analytical theory of linear emission of exchange spin waves from a Bloch domain wall, excited by a uniform microwave magnetic field.  The problem is reduced to a one-dimensional Schr\"odinger-like equation with a P\"oschl-Teller potential and a driving term of the same profile. The emission of plane spin waves is observed at excitation frequencies above a threshold value, as a result of a linear process. The height-to-width aspect ratio of the P\"oschl-Teller profile for a domain wall is found to correspond to a local maximum of the emission efficiency. Furthermore, for a tailored P\"oschl-Teller potential with a variable aspect ratio, particular values of the latter can lead to enhanced or even completely suppressed emission.

\end{abstract}

\maketitle


Wave generation is both an essential topic in wave physics and a prerequisite of any technology exploiting waves. Conventionally, waves are excited using an antenna, with their wavelength being limited by the antenna's size. Alternatively, we could use an inhomogeneity (either deliberately introduced, or naturally-occuring) in the medium and then apply a uniform, oscillatory external field to generate a wave. Small wavelength excitations require equally small antennas or inhomogeneities to generate them.

In magnonics \cite{kruglyak_magnonics_2010, krawczyk_review_2014, chumak_magnon_2015}, the study of spin waves, we are fortunate that inhomogeneities with nanoscale dimensions naturally occur in magnetic materials: domain walls. These inhomogeneities are the transition regions between domains of uniformly aligned magnetization, and can have dimensions down to a few nanometers, depending on the material. Domain walls have been studied in great detail, due to a number of interesting properties: their magnetic field and current-driven motion \cite{guslienko_dynamics_2008,yoshimura_soliton-like_2016}, their ability to channel spin waves \cite{garcia-sanchez_narrow_2015,wang_domain_2015,wagner_magnetic_2016}, and the unusual reflectionless behavior for spin waves passing through them \cite{thiele_excitation_1973}. Recently, there have also been numerical \cite{hermsdoerfer_spin-wave_2009, roy_micromagnetic_2010-1} and experimental \cite{mozooni_direct_2015,sluka_stacked_2015} reports of pinned domain walls \emph{generating} spin waves, with wavelengths down to tens of nanometers \cite{van_de_wiele_tunable_2016}. The origin of the observed spin wave emission has typically been attributed to the domain wall oscillations, generated by the applied microwave magnetic field \cite{hermsdoerfer_spin-wave_2009, roy_micromagnetic_2010-1,mozooni_direct_2015,sluka_stacked_2015} or spin-polarized current \cite{le_maho_spin-wave_2009,van_de_wiele_tunable_2016}. 

In this letter, we report an analytical theory that demonstrates emission of exchange spin waves from a Bloch domain wall driven by a uniform microwave magnetic field, as a result of a \emph{linear} process. The problem is reduced to that of the P\"oschl-Teller potential in a Schr\"odinger-like equation - an exactly solvable model, of particular interest in quantum mechanics \cite{flugge_practical_1971} and optics \cite{yildirim_intensity-dependent_2006,thekkekara_optical_2014}. This potential is mostly known for its peculiar property of 100\% transmission of incident waves at \emph{any} frequency, for certain parameters of the potential \cite{lekner_reflectionless_2007}. While forming such a potential in other systems is difficult, serendipitously the reflectionless P\"oschl-Teller potential \emph{exactly} describes the graded magnonic index profile \cite{davies_graded-index_2015} due to a Bloch domain wall, allowing the peculiar behavior to be both investigated and exploited in magnetic systems \cite{vasiliev_spin_2007-1}. Furthermore, when the domain wall is driven by a uniform microwave magnetic field, the P\"oschl-Teller profile happens to be present not only as the potential, but also as a driving term in the obtained Schr\"odinger-like equation. Strikingly, when we manipulate the aspect ratio of the profile from that of a domain wall, we reveal novel effects on the waves in our system, which are not present for the quantum-mechanical analogue (which has no driving term).
\begin{figure}
\centering
\includegraphics[width=\linewidth]{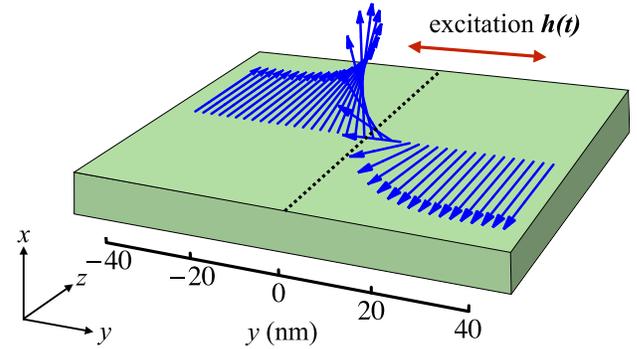}
\caption{(Color online) The studied system: a thin film with two antiparallel magnetic domains separated by a Bloch domain wall. The dotted line indicates the domain wall centre. The blue arrows represent the static magnetization configuration $\vec{M}_\text{S}$, with its magnitude $|\vec{M}_\text{S}|=M_0$ arbitrarily sized for clarity.}
\label{fig:fig_1_system}
\end{figure}

A thin film with infinite extent in the $y$-$z$ plane contains two antiparallel domains separated by a Bloch domain wall, as shown in Fig. \ref{fig:fig_1_system}. The magnetic energy density $W$ of this system is \cite{gurevich_magnetization_1996, landau_electrodynamics_1960}
\begin{align}
\label{eq:W}
W  = & \frac{\alpha}{2}(\partial_y\vec{M})^2 -\frac{\beta_{\parallel}}{2}(\vec{M}\cdot\hat{\vec{z}})^2 +\frac{\beta_{\perp}}{2}(\vec{M}\cdot\hat{\vec{x}})^2 -\vec{h}\cdot\vec{M},
\end{align}
\noindent where $\alpha$ is the exchange constant, $\beta_\parallel$ and $\beta_\perp$ are the constants of the easy axis and easy plane anisotropies respectively, $\hat{\vec{z}}$ and $\hat{\vec{x}}$ are unit vectors in the corresponding easy and hard magnetization directions, and $\vec{h}=h\exp(-i\omega t) \hat{\vec{y}}$ is the driving microwave magnetic field at frequency $\omega$. Minimizing the energy in spherical coordinates leads to the well-known profile of a Bloch domain wall \cite{landau_theory_1935-1}
\begin{align}
\label{eq:theta}
\theta& = 2\arctan\left[ \exp\left(y/\lambda_{B}\right) \right],
\end{align}
where $\theta$ is the polar angle and $\lambda_B=\sqrt{\alpha/(\beta_{\parallel} + \beta_{\perp})}$ is the domain wall width. The center of the domain wall is chosen to be at $y=0$. The dynamics of the system can be described by the Landau-Lifshitz equation
\begin{align}
\label{eq:LL}
\frac{\partial \vec{M}}{\partial t} & =  \gamma \left(\vec{M} \times \frac{\delta W}{\delta \vec{M}} \right),
\end{align}
where $\gamma$ is the gyromagnetic ratio, and for simplicity we ignore damping. The magnetization can be written as $\vec{M}=\vec{M}_{\text{S}}+\vec{m}$, where $\vec{m}$ is a small, time-dependent perturbation to the static magnetization $\vec{M}_{\text{S}}$. Treating $\vec{m}$ and $\vec{h}$ as small quantities, equations \eqref{eq:W} and \eqref{eq:LL} are linearized yeilding three coupled differential equations - one for each magnetisation component.                                                                                                                                                                                                                                                                       

It is convenient to carry out a coordinate transformation, where the new primed reference frame rotates around the $y$ axis (thus $y'=y$) and the new $z'$ axis is set to point along the static magnetization $\vec{M}_\text{S}$. In this frame, we are only concerned with precession in the $x'$-$y'$ plane. So, equation \eqref{eq:LL} reduces further to two coupled differential equations in $\vec{m}'= (m_x',
m_y')^T$, the time-dependent magnetization in the rotated frame,
\begin{align}
\label{eq:inhom}
\frac{1}{\alpha\gamma M_0} \frac{\partial}{\partial t} \vec{m}' = &
i \boldsymbol\sigma_y
\left[ \frac{\partial^2}{\partial y^2} -\frac{\beta_{\parallel}}{\alpha} - \beta(y) \right] 
 \vec{m}'
\nonumber \\
& + 
 \left( \begin{array}{c c}
0 & 0 \\
\frac{\beta_{\perp}}{\alpha} & 0 \end{array} \right) 
 \vec{m}'
  +
 \left( \begin{array}{c}
\frac{h}{\alpha} \\
0 \end{array} \right),
\end{align}
where $\boldsymbol\sigma_y$ is the Pauli spin matrix and $\beta(y) = -(2/\lambda_B^2)\,\text{sech}^2\left(y/\lambda_{B}\right)$. Following the approach from \cite{gorobets_excitation_1998}, we represent $\vec{m}'$ as a sum of two dynamic contributions due to the microwave excitation: (1) uniform background precession $\vec{m}_h'$, which is present irrespective of the presence of the domain wall in the sample; (2) non-uniform propagating spin waves $\vec{m}_\beta'$ excited due to the domain wall. Converting to the frequency domain and denoting the Fourier transformed variables with a tilde, we write the inhomogeneous equation in $\tilde{\vec{m}}_\beta' $ as 
\begin{align}
\label{eq:inhomFT}
 i \boldsymbol\sigma_y
  \beta(y) \tilde{\vec{m}}_h'  = & i \boldsymbol\sigma_y
 \left[ \partial_y^2 -\frac{\beta_{\parallel}}{\alpha} - \beta(y) \right] \tilde{\vec{m}}_\beta' \nonumber \\
& + 
 \left( \begin{array}{c c}
\frac{i\Omega}{\alpha}  & 0 \\
\frac{\beta_{\perp}}{\alpha} & \frac{i\Omega}{\alpha} \end{array} \right) \tilde{\vec{m}}_\beta',
\end{align}
where the uniform precession $\tilde{\vec{m}}_h'$ is 
\begin{align}
\label{eq:mh}
\tilde{\vec{m}}_h'  &= \frac{1}{-\Omega^2 + \beta_\parallel(\beta_\parallel + \beta_\perp)}
\left( \begin{array}{c}
-i\Omega \tilde{h}(\omega) \\
 (\beta_\parallel + \beta_{\perp})\tilde{h}(\omega) \end{array} \right),
\end{align}
and we have introduced the dimensionless frequency $\Omega=\omega/\gamma M_0$. There are some important features of equation \eqref{eq:inhomFT} to notice. Firstly, the term inside the square brackets resembles the Schr\"odinger equation with P\"oschl-Teller potential $\beta(y)$ \cite{wieser_quantized_2009, gonzalez_spin_2010, borys_spin-wave_2016}. Secondly, the driving term on the left hand side also contains $\beta(y)$, which will modify our results compared to the traditional Schr\"odinger equation with P\"oschl-Teller potential. Finally, the off-diagonal term $\beta_\perp/\alpha$, which only multiplies $m_{\beta,x}'$, will lead to ellipticity in the magnetization, as expected for a thin film. 

We separate the solution of the homogeneous equation corresponding to \eqref{eq:inhomFT} into a product  of a constant vector amplitude and a scalar function of position
\begin{align}
\label{eq:mg'}
 \vec{m}_{\beta,G}^{\pm'}=  & \left( \begin{array}{c}
 1 \\
 \pm i \sqrt{1-\frac{\beta_\perp}{\alpha\Lambda^\pm}}
 \end{array}\right) \varphi(y) 
 \equiv \vec{a}^\pm \varphi(y). 
\end{align}
The factor $\varphi(y)$ is the well-known solution to the Schr\"odinger equation with P\"oschl-Teller potential well \cite{flugge_practical_1971,gorobets_excitation_1998}, with eigenvalues $\Lambda^\pm$:
\begin{equation}
\label{eq:lambda}
\Lambda^\pm = \frac{1}{\alpha} \left(\frac{\beta_\perp}{2} \pm \xi \right), \quad \xi= \sqrt{\frac{\beta_\perp^2}{4} + \Omega^2}.
\end{equation}
We write $\varphi(y)=C_1^\pm w_1^\pm + C_2^\pm w_2^\pm$,
where
\begin{widetext}
\begin{align}
\label{eq:w1}
w_1^\pm = & [\text{cosh}(y/\lambda_B)]^{ik^\pm\lambda_B} F\left[-ik^\pm\lambda_B -1, -ik^\pm\lambda_B+2, -ik^\pm\lambda_B+1; \zeta \right], 
\\ \nonumber \\
w_2^\pm = & \frac{\Gamma(-ik^\pm \lambda_B+1)\Gamma(-ik^\pm \lambda_B)}{\Gamma(-ik^\pm \lambda_B-1)\Gamma(-ik^\pm \lambda_B+2)} \, \left [ \text{cosh}(y/\lambda_B)\zeta \right ] ^{ik^\pm\lambda_B} 
F\left[-1,2,ik^\pm\lambda_B+1;\zeta \right] 
\nonumber \\ \nonumber \\
\label{eq:w2}
& -   \frac{\Gamma(ik^\pm \lambda_B+1)\Gamma(-ik^\pm \lambda_B)}{\Gamma(-1)\Gamma(2)} \, \left [\text{cosh}(y/\lambda_B) \right ]^{ik^\pm\lambda_B}F\left[-ik^\pm \lambda_B-1, -ik^\pm \lambda_B+2, -ik^\pm \lambda_B+1; \zeta \right].
\end{align}
\end{widetext}
$F[a,b,c;\zeta]$ is the hypergeometric function with $\zeta=[1-\text{tanh}(y/\lambda_B)]/2$, the wave numbers are $k^\pm =\sqrt{-\Lambda^\pm- (\beta_\parallel/\alpha)}$, and $C_1^\pm$ and $C_2^\pm$ are constants to be found using the boundary conditions. We can see from our definition of $\Lambda^\pm$ in \eqref{eq:lambda} that only the $k^-$ solution can be real-valued, and only when $\Omega > \sqrt{ \beta_\parallel(\beta_\parallel + \beta_\perp)}$. We find that the $k^+$ solution does affect the form of the dynamic magnetization within the domain wall, and so we retain both solutions.

Now, we can use the method of variation of parameters to find the solution to the inhomogeneous equation, as
%
\begin{align}
\label{eq:mb}
\tilde{\vec{m}}_\beta'  = &\sum_{\sigma=+,-} \vec{a}^\sigma\Bigg \{ C_1^\sigma w_1^\sigma + C_2^\sigma w_2^\sigma  
+  w_1^\sigma f_1^\sigma(0,y) \nonumber \\
& - w_2 ^\sigma f_2^\sigma (0,y)\Bigg \}, 
\end{align}
with
\begin{align}
f_{1(2)}^\pm(p,q) &= -\frac{NA^\pm}{W^\pm}  \int_p^q  \beta(y)  w_{2(1)}^\pm \; \text{d} y', \nonumber \\
 N  &=
- \frac{ i \Omega \tilde{h}(\omega) }{2\xi [-\Omega^2 + \beta_\parallel(\beta_\parallel + \beta_\perp)]},  \nonumber \\
 A^\pm &=  \pm \left ( \frac{\beta_{\perp}}{2} + \beta_\parallel \right) - \xi, \nonumber \\
  W^\pm &= \frac{ik^\pm}{2^{i k^\pm \lambda_B}} \frac{\Gamma(-ik^\pm \lambda_B+1)\Gamma(-ik^\pm \lambda_B)}{\Gamma(-ik^\pm \lambda_B-1)\Gamma(-ik^\pm \lambda_B+2)}. 
  \nonumber
\end{align}
%
$W^+$ and $W^-$ are the Wronskians for the $k^+$ and $k^-$ solutions, respectively \cite{gorobets_excitation_1998}. 

Finally, to find $C_1^\pm$ and $C_2^\pm$, we require that the magnetization $\tilde{\vec{m}}_\beta'$ tends to plane waves that propagate outwards from the domain wall, i.e., $ \lim_{y\to\pm\infty} \{\tilde{\vec{m}}_{\beta}' \} \approx \vec{S}^+(\omega)e^{\pm ik^+ y} + \vec{S}^-(\omega)e^{\pm ik^- y}$. As before, we are using the convention that superscripts relate to the two wave number solutions, while here the $\pm$ sign in the exponentials relates to the $\pm \infty$ limits. So, we find 
 \begin{subequations}
 \begin{align}
 C_{1(2)}^\pm & = f_{1(2)}^\pm \big(-\infty(0), 0 (+\infty) \big), \\
 \label{eq:Sw1}
 \vec{S}^\pm(\omega)  & = \frac{\vec{a}^\pm}{ 2^{i k^\pm \lambda_B}} f_1^\pm(-\infty, +\infty).
  \end{align}
 \end{subequations}
We now have the full solution for the magnetization $\tilde{\vec{m}}_\beta'$, and therefore know the spin wave amplitude at the asymptotic limits $\vec{S}^\pm(\omega)$. However, since $k^+$ is always imaginary, we are only concerned with the $\vec{S}^-(\omega)$ solution. 

We study the solutions obtained for a permalloy-like sample \footnote{We use the following parameters throughout this paper: $M_0$ = 800 erg$\cdot$G$^{-1}\cdot$cm$^{-3}$, $\gamma = 1.76\times 10^7$ Hz$\cdot$Oe$^{-1}$, $\alpha = 3.125 \times10^{-12}$  cm$^{2}$, $\beta_\parallel = 0.1$ ($K=32\times10^3$erg$\cdot$cm$^{-3}$), $\beta_\perp = 10$ ($K=32\times10^5$erg$\cdot$cm$^{-3}$)} - the corresponding width of the Bloch domain wall is $\lambda_B=6$nm. Let us begin by considering the character of the magnetization dynamics near the domain wall. To visualize the dynamics, we convert our variables back into the laboratory frame and plot them as arrows.  The results are shown in Fig. \ref{fig:fig_2_precession}, and the full animation from different viewpoints is provided in \footnote{See Supplemental Material at [link TBC]}.  We can see that the magnetization precession in the domain wall has a larger amplitude than in the adjacent domains. Moreover, the domain wall center, i.e., the position of the magnetization with the largest $x$ component, appears to move back and forth along the $y$ direction. 
\begin{figure}
\centering
\includegraphics[width=\linewidth]{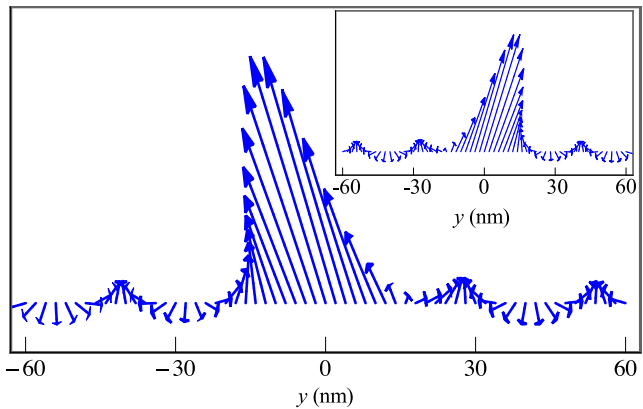}
\caption{(Color online) Projection of the magnetization vectors onto the $x-y$ plane, in the unrotated frame, for  phase $=0$ and (inset) phase $=\pi$. The $y$ position and orientation of the vectors are accurate, while the static magnetization length has been arbitrarily reduced for clarity.}
\label{fig:fig_2_precession}
\end{figure}

The apparent domain wall motion is not however the source of the emitted spin waves. Rather, this motion is the small amplitude precession given by the solution of the linearized Landau-Lifshitz equation, and is of the same order as the emitted spin waves. Indeed, the precessional modes resulting from a linear theory must obey the superposition principle, while the full non-linear Landau-Lifshitz equation would need to be solved to account for any interaction between different precessional modes. Examples of a non-linear generation of spin waves from a domain wall can be found in, e.g., Refs. \cite{hermsdoerfer_spin-wave_2009,boone_magnetic_2010} where the domain wall oscillations at frequency $\omega$ were observed to emit spin waves at twice the frequency, i.e., $2\omega$. Our theory suggests that the spin wave emission from domain walls, at a frequency equal to that of the driving magnetic field \cite{mozooni_direct_2015,sluka_stacked_2015} or spin-polarized current \cite{van_de_wiele_tunable_2016} should rather be interpreted as a linear excitation due to the magnetic inhomogeneity \cite{schlomann_generation_1964-1} (``graded magnonic index" \cite{davies_graded-index_2015}) created by the domain wall, when excited by a uniform magnetic field. 

\begin{figure}
\centering
\includegraphics[width=\linewidth]{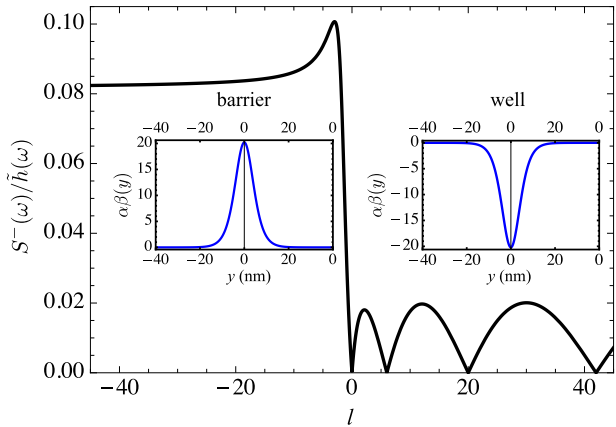}
\caption{(Color online) Spin wave amplitude vs. $l$ for potential barrier ($l<0$) and well ($l>0$) for frequency $f=50$GHz, with shape of the potentials shown for $l=-2$ for the barrier (left inset) and $l=2$ for the well (right inset).}
\label{fig:fig_3_height}
\end{figure}

Through analysing the behavior of the P\"oschl-Teller potential with different parameters, we discover that it is actually non-trivial that a domain wall emits spin waves. To elucidate this, we refer to the theory describing the P\"oschl-Teller potential well for \emph{incident} waves \cite{lekner_reflectionless_2007} - it is well-known to have `special' values of height at which it becomes reflectionless. If we write the profile as
\begin{equation}
\beta_l(y)= - \frac{l}{\lambda_B^2} \text{sech}^2\left(\frac{y}{\lambda_B} \right),  \quad l=n(n+1),
\end{equation}
the profiles that are reflectionless for incident waves can be identified as those with integer $n$. We now use \eqref{eq:Sw1} to investigate how changing $l$ affects the \emph{emission} of spin waves from the profile. In Fig. \ref{fig:fig_3_height} we sweep $l$ from negative values (which represent a potential barrier - there are no solutions for $n$ in this case) through to positive values (potential well, which has corresponding values of $n$). We observe for the potential well that at certain values of $l$, which correspond to \emph{even} $n$, the spin wave emission is zero, and the spin waves are confined within the domain wall region. Furthermore, profiles with odd $n$ are local maxima. So, the presence of sech$^2(y/\lambda_B)$ in both the potential and the driving term in \eqref{eq:inhomFT} leads to a different set of `special' values, corresponding to either strong wave emission or its complete suppression. The particular value of $l=2$ ($n=1$) for a domain wall happens to correspond to a local maximum condition for spin wave emission. However, a potential barrier of any height generates spin waves much more efficiently than the potential well solutions. The peak at around $l=-3$ generates spin waves most efficiently compared to any other profile height, although this optimal value depends on the frequency for a given set of the other parameters. 

\begin{figure}
\centering
\includegraphics[width=\linewidth]{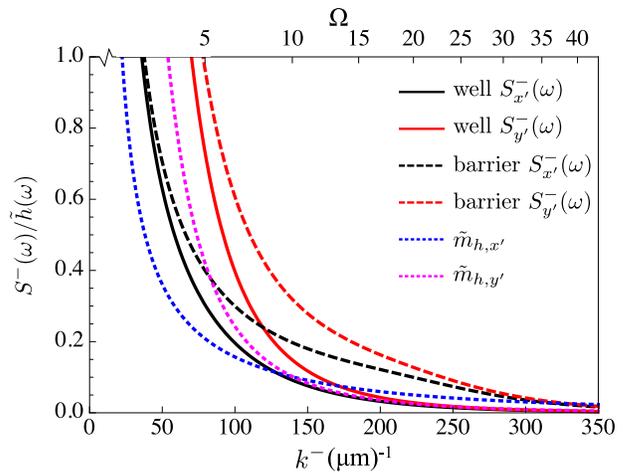}
\caption{(Color online) Amplitude of the spin waves generated by a P\"oschl-Teller potential well (solid lines) and potential barrier (dashed lines), compared to the amplitude of uniform precession induced by the external field (dotted lines), showing $x'$ and $y'$ components (colors indicated on the graph). All quantities are normalised by the external field $\tilde{h}(\omega)$. $\vec{S}^-(\omega)$ is a function of $k^-$ (bottom axis) and thus $\Omega$ (top axis), and $\tilde{\vec{m}}_{h}$ is only a function of $\Omega$.}
\label{fig:fig_4_SW}
\end{figure}

Fig. \ref{fig:fig_4_SW} compares the frequency dependence of the $x'$ and $y'$ components of the spin wave amplitude $\vec{S}^-(\omega)$, for both a P\"oschl-Teller potential well (domain wall) and barrier. For comparison, we include the magnitude of the uniform precession $\tilde{\vec{m}}_h$, excited by the same field. 
The difference between $S_{x'}^-(\omega)$ and $S_{y'}^-(\omega)$ at small wave numbers shows that the precession is elliptical, with the ellipticity decreasing with increasing frequency. The frequency dependences of the precession amplitude far from the domain wall are different for propagating spin waves $\vec{S}^-(\omega)$ and the uniform precession $\tilde{\vec{m}}_h'$. As a result, for a domain wall (more generally, a P\"oschl-Teller \emph{well} profile) the out of plane component $S^-_{x'}(\omega)$ is only larger in amplitude than $\tilde{\vec{m}}_h'$ at low frequencies, i.e., for $k^-\lesssim$ 150($\mu$m)$^{-1}$. However, for a P\"oschl-Teller \emph{barrier} profile, $S^-_{x'}(\omega)$ exceeds $\tilde{\vec{m}}_h'$ up to much higher frequencies, i.e., for $k^-\lesssim$ 300($\mu$m)$^{-1}$. This shows that, despite domain walls being such efficient magnonic emitters, an even better efficiency could be achieved by tailoring the local effective magnetic field (through modification of e.g. the anisotropy strength \cite{li_microstructures:_2002,vasiliev_spin_2007-1}) to form a P\"oschl-Teller potential barrier instead.  

In summary, we have investigated theoretically the origin and behavior of exchange spin waves generated by a Bloch domain wall.  We find that the graded magnonic index in the domain wall region leads to emission of spin waves at the frequency of the driving uniform harmonic microwave field, and with amplitude scaling linearly with the field strength. The identified linear character of the spin wave emission process excludes its interpretation as resulting from the domain wall motion. The depth of the P\"oschl-Teller profile due to the domain wall is naturally sized so as to maximize the spin wave emission, while we also find that certain `heights' of the profile can lead to spin wave confinement within, rather than emission from, the profile. Further enhancement of the spin wave emission could be achieved using nanofabrication to artificially shape the graded magnonic index, to form a P\"oschl-Teller barrier rather than a well.  

\begin{acknowledgments}
This research has received funding from the Engineering and Physical Sciences Research Council (EPSRC) of the United Kingdom, via the EPSRC center for Doctoral Training in Metamaterials (Grant No. EP/L015331/1), and the European Union's Horizon 2020 research and innovation program under Marie Sk\l{}odowska-Curie Grant Agreement No. 644348 (MagIC). SARH would like to thank the Royal Society and TATA for financial support.
\end{acknowledgments}


%



\begin{thebibliography}{33}%
\makeatletter
\providecommand \@ifxundefined [1]{%
 \@ifx{#1\undefined}
}%
\providecommand \@ifnum [1]{%
 \ifnum #1\expandafter \@firstoftwo
 \else \expandafter \@secondoftwo
 \fi
}%
\providecommand \@ifx [1]{%
 \ifx #1\expandafter \@firstoftwo
 \else \expandafter \@secondoftwo
 \fi
}%
\providecommand \natexlab [1]{#1}%
\providecommand \enquote  [1]{``#1''}%
\providecommand \bibnamefont  [1]{#1}%
\providecommand \bibfnamefont [1]{#1}%
\providecommand \citenamefont [1]{#1}%
\providecommand \href@noop [0]{\@secondoftwo}%
\providecommand \href [0]{\begingroup \@sanitize@url \@href}%
\providecommand \@href[1]{\@@startlink{#1}\@@href}%
\providecommand \@@href[1]{\endgroup#1\@@endlink}%
\providecommand \@sanitize@url [0]{\catcode `\\12\catcode `\$12\catcode
  `\&12\catcode `\#12\catcode `\^12\catcode `\_12\catcode `\%12\relax}%
\providecommand \@@startlink[1]{}%
\providecommand \@@endlink[0]{}%
\providecommand \url  [0]{\begingroup\@sanitize@url \@url }%
\providecommand \@url [1]{\endgroup\@href {#1}{\urlprefix }}%
\providecommand \urlprefix  [0]{URL }%
\providecommand \Eprint [0]{\href }%
\providecommand \doibase [0]{http://dx.doi.org/}%
\providecommand \selectlanguage [0]{\@gobble}%
\providecommand \bibinfo  [0]{\@secondoftwo}%
\providecommand \bibfield  [0]{\@secondoftwo}%
\providecommand \translation [1]{[#1]}%
\providecommand \BibitemOpen [0]{}%
\providecommand \bibitemStop [0]{}%
\providecommand \bibitemNoStop [0]{.\EOS\space}%
\providecommand \EOS [0]{\spacefactor3000\relax}%
\providecommand \BibitemShut  [1]{\csname bibitem#1\endcsname}%
\let\auto@bib@innerbib\@empty
\bibitem [{\citenamefont {Kruglyak}\ \emph {et~al.}(2010)\citenamefont
  {Kruglyak}, \citenamefont {Demokritov},\ and\ \citenamefont
  {Grundler}}]{kruglyak_magnonics_2010}%
  \BibitemOpen
  \bibfield  {author} {\bibinfo {author} {\bibfnamefont {V.~V.}\ \bibnamefont
  {Kruglyak}}, \bibinfo {author} {\bibfnamefont {S.~O.}\ \bibnamefont
  {Demokritov}}, \ and\ \bibinfo {author} {\bibfnamefont {D.}~\bibnamefont
  {Grundler}},\ }\href {\doibase 10.1088/0022-3727/43/26/264001} {\bibfield
  {journal} {\bibinfo  {journal} {J. Phys. D: Appl. Phys.}\ }\textbf {\bibinfo
  {volume} {43}},\ \bibinfo {pages} {264001} (\bibinfo {year}
  {2010})}\BibitemShut {NoStop}%
\bibitem [{\citenamefont {Krawczyk}\ and\ \citenamefont
  {Grundler}(2014)}]{krawczyk_review_2014}%
  \BibitemOpen
  \bibfield  {author} {\bibinfo {author} {\bibfnamefont {M.}~\bibnamefont
  {Krawczyk}}\ and\ \bibinfo {author} {\bibfnamefont {D.}~\bibnamefont
  {Grundler}},\ }\href {\doibase 10.1088/0953-8984/26/12/123202} {\bibfield
  {journal} {\bibinfo  {journal} {J. Phys.: Condens. Matter}\ }\textbf
  {\bibinfo {volume} {26}},\ \bibinfo {pages} {123202} (\bibinfo {year}
  {2014})}\BibitemShut {NoStop}%
\bibitem [{\citenamefont {Chumak}\ \emph {et~al.}(2015)\citenamefont {Chumak},
  \citenamefont {Vasyuchka}, \citenamefont {Serga},\ and\ \citenamefont
  {Hillebrands}}]{chumak_magnon_2015}%
  \BibitemOpen
  \bibfield  {author} {\bibinfo {author} {\bibfnamefont {A.~V.}\ \bibnamefont
  {Chumak}}, \bibinfo {author} {\bibfnamefont {V.~I.}\ \bibnamefont
  {Vasyuchka}}, \bibinfo {author} {\bibfnamefont {A.~A.}\ \bibnamefont
  {Serga}}, \ and\ \bibinfo {author} {\bibfnamefont {B.}~\bibnamefont
  {Hillebrands}},\ }\href {\doibase 10.1038/nphys3347} {\bibfield  {journal}
  {\bibinfo  {journal} {Nat. Phys.}\ }\textbf {\bibinfo {volume} {11}},\
  \bibinfo {pages} {453} (\bibinfo {year} {2015})}\BibitemShut {NoStop}%
\bibitem [{\citenamefont {Guslienko}\ \emph {et~al.}(2008)\citenamefont
  {Guslienko}, \citenamefont {Lee},\ and\ \citenamefont
  {Kim}}]{guslienko_dynamics_2008}%
  \BibitemOpen
  \bibfield  {author} {\bibinfo {author} {\bibfnamefont {K.~Y.}\ \bibnamefont
  {Guslienko}}, \bibinfo {author} {\bibfnamefont {J.~Y.}\ \bibnamefont {Lee}},
  \ and\ \bibinfo {author} {\bibfnamefont {S.~K.}\ \bibnamefont {Kim}},\ }\href
  {\doibase 10.1109/TMAG.2008.2002775} {\bibfield  {journal} {\bibinfo
  {journal} {IEEE Trans. Magn.}\ }\textbf {\bibinfo {volume} {44}},\ \bibinfo
  {pages} {3079} (\bibinfo {year} {2008})}\BibitemShut {NoStop}%
\bibitem [{\citenamefont {Yoshimura}\ \emph {et~al.}(2016)\citenamefont
  {Yoshimura}, \citenamefont {Kim}, \citenamefont {Taniguchi}, \citenamefont
  {Tono}, \citenamefont {Ueda}, \citenamefont {Hiramatsu}, \citenamefont
  {Moriyama}, \citenamefont {Yamada}, \citenamefont {Nakatani},\ and\
  \citenamefont {Ono}}]{yoshimura_soliton-like_2016}%
  \BibitemOpen
  \bibfield  {author} {\bibinfo {author} {\bibfnamefont {Y.}~\bibnamefont
  {Yoshimura}}, \bibinfo {author} {\bibfnamefont {K.-J.}\ \bibnamefont {Kim}},
  \bibinfo {author} {\bibfnamefont {T.}~\bibnamefont {Taniguchi}}, \bibinfo
  {author} {\bibfnamefont {T.}~\bibnamefont {Tono}}, \bibinfo {author}
  {\bibfnamefont {K.}~\bibnamefont {Ueda}}, \bibinfo {author} {\bibfnamefont
  {R.}~\bibnamefont {Hiramatsu}}, \bibinfo {author} {\bibfnamefont
  {T.}~\bibnamefont {Moriyama}}, \bibinfo {author} {\bibfnamefont
  {K.}~\bibnamefont {Yamada}}, \bibinfo {author} {\bibfnamefont
  {Y.}~\bibnamefont {Nakatani}}, \ and\ \bibinfo {author} {\bibfnamefont
  {T.}~\bibnamefont {Ono}},\ }\href {\doibase 10.1038/nphys3535} {\bibfield
  {journal} {\bibinfo  {journal} {Nat. Phys.}\ }\textbf {\bibinfo {volume}
  {12}},\ \bibinfo {pages} {157} (\bibinfo {year} {2016})}\BibitemShut
  {NoStop}%
\bibitem [{\citenamefont {Garcia-Sanchez}\ \emph {et~al.}(2015)\citenamefont
  {Garcia-Sanchez}, \citenamefont {Borys}, \citenamefont {Soucaille},
  \citenamefont {Adam}, \citenamefont {Stamps},\ and\ \citenamefont
  {Kim}}]{garcia-sanchez_narrow_2015}%
  \BibitemOpen
  \bibfield  {author} {\bibinfo {author} {\bibfnamefont {F.}~\bibnamefont
  {Garcia-Sanchez}}, \bibinfo {author} {\bibfnamefont {P.}~\bibnamefont
  {Borys}}, \bibinfo {author} {\bibfnamefont {R.}~\bibnamefont {Soucaille}},
  \bibinfo {author} {\bibfnamefont {J.-P.}\ \bibnamefont {Adam}}, \bibinfo
  {author} {\bibfnamefont {R.~L.}\ \bibnamefont {Stamps}}, \ and\ \bibinfo
  {author} {\bibfnamefont {J.-V.}\ \bibnamefont {Kim}},\ }\href {\doibase
  10.1103/PhysRevLett.114.247206} {\bibfield  {journal} {\bibinfo  {journal}
  {Phys. Rev. Lett.}\ }\textbf {\bibinfo {volume} {114}},\ \bibinfo {pages}
  {247206} (\bibinfo {year} {2015})}\BibitemShut {NoStop}%
\bibitem [{\citenamefont {Wang}\ and\ \citenamefont
  {Wang}(2015)}]{wang_domain_2015}%
  \BibitemOpen
  \bibfield  {author} {\bibinfo {author} {\bibfnamefont {X.~S.}\ \bibnamefont
  {Wang}}\ and\ \bibinfo {author} {\bibfnamefont {X.~R.}\ \bibnamefont
  {Wang}},\ }\href@noop {} {\bibfield  {journal} {\bibinfo  {journal}
  {arXiv:1512.05965v2 [cond-mat.mes-hall]}\ } (\bibinfo {year}
  {2015})}\BibitemShut {NoStop}%
\bibitem [{\citenamefont {Wagner}\ \emph {et~al.}(2016)\citenamefont {Wagner},
  \citenamefont {K{\'a}kay}, \citenamefont {Schultheiss}, \citenamefont
  {Henschke}, \citenamefont {Sebastian},\ and\ \citenamefont
  {Schultheiss}}]{wagner_magnetic_2016}%
  \BibitemOpen
  \bibfield  {author} {\bibinfo {author} {\bibfnamefont {K.}~\bibnamefont
  {Wagner}}, \bibinfo {author} {\bibfnamefont {A.}~\bibnamefont {K{\'a}kay}},
  \bibinfo {author} {\bibfnamefont {K.}~\bibnamefont {Schultheiss}}, \bibinfo
  {author} {\bibfnamefont {A.}~\bibnamefont {Henschke}}, \bibinfo {author}
  {\bibfnamefont {T.}~\bibnamefont {Sebastian}}, \ and\ \bibinfo {author}
  {\bibfnamefont {H.}~\bibnamefont {Schultheiss}},\ }\href {\doibase
  10.1038/nnano.2015.339} {\bibfield  {journal} {\bibinfo  {journal} {Nat.
  Nanotechnol.}\ }\textbf {\bibinfo {volume} {11}},\ \bibinfo {pages} {432}
  (\bibinfo {year} {2016})}\BibitemShut {NoStop}%
\bibitem [{\citenamefont {Thiele}(1973)}]{thiele_excitation_1973}%
  \BibitemOpen
  \bibfield  {author} {\bibinfo {author} {\bibfnamefont {A.~A.}\ \bibnamefont
  {Thiele}},\ }\href {\doibase https://doi.org/10.1103/PhysRevB.7.391}
  {\bibfield  {journal} {\bibinfo  {journal} {Phys. Rev. B}\ }\textbf {\bibinfo
  {volume} {7}},\ \bibinfo {pages} {391} (\bibinfo {year} {1973})}\BibitemShut
  {NoStop}%
\bibitem [{\citenamefont {Hermsdoerfer}\ \emph {et~al.}(2009)\citenamefont
  {Hermsdoerfer}, \citenamefont {Schultheiss}, \citenamefont {Rausch},
  \citenamefont {Sch{\"a}fer}, \citenamefont {Leven}, \citenamefont {Kim},\
  and\ \citenamefont {Hillebrands}}]{hermsdoerfer_spin-wave_2009}%
  \BibitemOpen
  \bibfield  {author} {\bibinfo {author} {\bibfnamefont {S.~J.}\ \bibnamefont
  {Hermsdoerfer}}, \bibinfo {author} {\bibfnamefont {H.}~\bibnamefont
  {Schultheiss}}, \bibinfo {author} {\bibfnamefont {C.}~\bibnamefont {Rausch}},
  \bibinfo {author} {\bibfnamefont {S.}~\bibnamefont {Sch{\"a}fer}}, \bibinfo
  {author} {\bibfnamefont {B.}~\bibnamefont {Leven}}, \bibinfo {author}
  {\bibfnamefont {S.-K.}\ \bibnamefont {Kim}}, \ and\ \bibinfo {author}
  {\bibfnamefont {B.}~\bibnamefont {Hillebrands}},\ }\href {\doibase
  10.1063/1.3143225} {\bibfield  {journal} {\bibinfo  {journal} {Appl. Phys.
  Lett.}\ }\textbf {\bibinfo {volume} {94}},\ \bibinfo {pages} {223510}
  (\bibinfo {year} {2009})}\BibitemShut {NoStop}%
\bibitem [{\citenamefont {Roy}\ \emph {et~al.}(2010)\citenamefont {Roy},
  \citenamefont {Trypiniotis},\ and\ \citenamefont
  {Barnes}}]{roy_micromagnetic_2010-1}%
  \BibitemOpen
  \bibfield  {author} {\bibinfo {author} {\bibfnamefont {P.~E.}\ \bibnamefont
  {Roy}}, \bibinfo {author} {\bibfnamefont {T.}~\bibnamefont {Trypiniotis}}, \
  and\ \bibinfo {author} {\bibfnamefont {C.~H.~W.}\ \bibnamefont {Barnes}},\
  }\href {\doibase 10.1103/PhysRevB.82.134411} {\bibfield  {journal} {\bibinfo
  {journal} {Phys. Rev. B}\ }\textbf {\bibinfo {volume} {82}},\ \bibinfo
  {pages} {134411} (\bibinfo {year} {2010})}\BibitemShut {NoStop}%
\bibitem [{\citenamefont {Mozooni}\ and\ \citenamefont
  {McCord}(2015)}]{mozooni_direct_2015}%
  \BibitemOpen
  \bibfield  {author} {\bibinfo {author} {\bibfnamefont {B.}~\bibnamefont
  {Mozooni}}\ and\ \bibinfo {author} {\bibfnamefont {J.}~\bibnamefont
  {McCord}},\ }\href {\doibase 10.1063/1.4927598} {\bibfield  {journal}
  {\bibinfo  {journal} {Appl. Phys. Lett.}\ }\textbf {\bibinfo {volume}
  {107}},\ \bibinfo {pages} {042402} (\bibinfo {year} {2015})}\BibitemShut
  {NoStop}%
\bibitem [{\citenamefont {Sluka}\ \emph {et~al.}(2015)\citenamefont {Sluka},
  \citenamefont {Weigand}, \citenamefont {Kakay}, \citenamefont {Erbe},
  \citenamefont {Tyberkevych}, \citenamefont {Slavin}, \citenamefont {Deac},
  \citenamefont {Lindner}, \citenamefont {Fassbender}, \citenamefont {Raabe},\
  and\ \citenamefont {Wintz}}]{sluka_stacked_2015}%
  \BibitemOpen
  \bibfield  {author} {\bibinfo {author} {\bibfnamefont {V.}~\bibnamefont
  {Sluka}}, \bibinfo {author} {\bibfnamefont {M.}~\bibnamefont {Weigand}},
  \bibinfo {author} {\bibfnamefont {A.}~\bibnamefont {Kakay}}, \bibinfo
  {author} {\bibfnamefont {A.}~\bibnamefont {Erbe}}, \bibinfo {author}
  {\bibfnamefont {V.}~\bibnamefont {Tyberkevych}}, \bibinfo {author}
  {\bibfnamefont {A.}~\bibnamefont {Slavin}}, \bibinfo {author} {\bibfnamefont
  {A.}~\bibnamefont {Deac}}, \bibinfo {author} {\bibfnamefont {J.}~\bibnamefont
  {Lindner}}, \bibinfo {author} {\bibfnamefont {J.}~\bibnamefont {Fassbender}},
  \bibinfo {author} {\bibfnamefont {J.}~\bibnamefont {Raabe}}, \ and\ \bibinfo
  {author} {\bibfnamefont {S.}~\bibnamefont {Wintz}},\ }in\ \href {\doibase
  10.1109/INTMAG.2015.7157029} {\emph {\bibinfo {booktitle} {2015 {{IEEE
  Magnetics Conference}} ({{INTERMAG}})}}}\ (\bibinfo {year}
  {2015})\BibitemShut {NoStop}%
\bibitem [{\citenamefont {{Van de Wiele}}\ \emph {et~al.}(2016)\citenamefont
  {{Van de Wiele}}, \citenamefont {H{\"a}m{\"a}l{\"a}inen}, \citenamefont
  {Bal{\'a}{\v z}}, \citenamefont {Montoncello},\ and\ \citenamefont {{van
  Dijken}}}]{van_de_wiele_tunable_2016}%
  \BibitemOpen
  \bibfield  {author} {\bibinfo {author} {\bibfnamefont {B.}~\bibnamefont {{Van
  de Wiele}}}, \bibinfo {author} {\bibfnamefont {S.~J.}\ \bibnamefont
  {H{\"a}m{\"a}l{\"a}inen}}, \bibinfo {author} {\bibfnamefont {P.}~\bibnamefont
  {Bal{\'a}{\v z}}}, \bibinfo {author} {\bibfnamefont {F.}~\bibnamefont
  {Montoncello}}, \ and\ \bibinfo {author} {\bibfnamefont {S.}~\bibnamefont
  {{van Dijken}}},\ }\href {\doibase 10.1038/srep21330} {\bibfield  {journal}
  {\bibinfo  {journal} {Sci. Rep.}\ }\textbf {\bibinfo {volume} {6}},\ \bibinfo
  {pages} {21330} (\bibinfo {year} {2016})}\BibitemShut {NoStop}%
\bibitem [{\citenamefont {Le~Maho}\ \emph {et~al.}(2009)\citenamefont
  {Le~Maho}, \citenamefont {Kim},\ and\ \citenamefont
  {Tatara}}]{le_maho_spin-wave_2009}%
  \BibitemOpen
  \bibfield  {author} {\bibinfo {author} {\bibfnamefont {Y.}~\bibnamefont
  {Le~Maho}}, \bibinfo {author} {\bibfnamefont {J.-V.}\ \bibnamefont {Kim}}, \
  and\ \bibinfo {author} {\bibfnamefont {G.}~\bibnamefont {Tatara}},\ }\href
  {\doibase 10.1103/PhysRevB.79.174404} {\bibfield  {journal} {\bibinfo
  {journal} {Phys. Rev. B}\ }\textbf {\bibinfo {volume} {79}},\ \bibinfo
  {pages} {174404} (\bibinfo {year} {2009})}\BibitemShut {NoStop}%
\bibitem [{\citenamefont {Fl{\"u}gge}(1971)}]{flugge_practical_1971}%
  \BibitemOpen
  \bibfield  {author} {\bibinfo {author} {\bibfnamefont {S.}~\bibnamefont
  {Fl{\"u}gge}},\ }\href@noop {} {\emph {\bibinfo {title} {Practical {{Quantum
  Mechanics}}}}}\ (\bibinfo  {publisher} {{Springer}},\ \bibinfo {address}
  {Berlin, Heidelberg},\ \bibinfo {year} {1971})\BibitemShut {NoStop}%
\bibitem [{\citenamefont {Yildirim}\ and\ \citenamefont
  {Tomak}(2006)}]{yildirim_intensity-dependent_2006}%
  \BibitemOpen
  \bibfield  {author} {\bibinfo {author} {\bibfnamefont {H.}~\bibnamefont
  {Yildirim}}\ and\ \bibinfo {author} {\bibfnamefont {M.}~\bibnamefont
  {Tomak}},\ }\href {\doibase 10.1063/1.2194124} {\bibfield  {journal}
  {\bibinfo  {journal} {J. Appl. Phys.}\ }\textbf {\bibinfo {volume} {99}},\
  \bibinfo {pages} {093103} (\bibinfo {year} {2006})}\BibitemShut {NoStop}%
\bibitem [{\citenamefont {Thekkekara}\ \emph {et~al.}(2014)\citenamefont
  {Thekkekara}, \citenamefont {Achanta},\ and\ \citenamefont
  {Gupta}}]{thekkekara_optical_2014}%
  \BibitemOpen
  \bibfield  {author} {\bibinfo {author} {\bibfnamefont {L.~V.}\ \bibnamefont
  {Thekkekara}}, \bibinfo {author} {\bibfnamefont {V.~G.}\ \bibnamefont
  {Achanta}}, \ and\ \bibinfo {author} {\bibfnamefont {S.~D.}\ \bibnamefont
  {Gupta}},\ }\href {\doibase 10.1364/OE.22.017382} {\bibfield  {journal}
  {\bibinfo  {journal} {Opt. Express}\ }\textbf {\bibinfo {volume} {22}},\
  \bibinfo {pages} {17382} (\bibinfo {year} {2014})}\BibitemShut {NoStop}%
\bibitem [{\citenamefont {Lekner}(2007)}]{lekner_reflectionless_2007}%
  \BibitemOpen
  \bibfield  {author} {\bibinfo {author} {\bibfnamefont {J.}~\bibnamefont
  {Lekner}},\ }\href {\doibase 10.1119/1.2787015} {\bibfield  {journal}
  {\bibinfo  {journal} {Am. J. Phys.}\ }\textbf {\bibinfo {volume} {75}},\
  \bibinfo {pages} {1151} (\bibinfo {year} {2007})}\BibitemShut {NoStop}%
\bibitem [{\citenamefont {Davies}\ and\ \citenamefont
  {Kruglyak}(2015)}]{davies_graded-index_2015}%
  \BibitemOpen
  \bibfield  {author} {\bibinfo {author} {\bibfnamefont {C.~S.}\ \bibnamefont
  {Davies}}\ and\ \bibinfo {author} {\bibfnamefont {V.~V.}\ \bibnamefont
  {Kruglyak}},\ }\href {\doibase 10.1063/1.4932349} {\bibfield  {journal}
  {\bibinfo  {journal} {Low Temp. Phys.}\ }\textbf {\bibinfo {volume} {41}},\
  \bibinfo {pages} {760} (\bibinfo {year} {2015})}\BibitemShut {NoStop}%
\bibitem [{\citenamefont {Vasiliev}\ \emph {et~al.}(2007)\citenamefont
  {Vasiliev}, \citenamefont {Kruglyak}, \citenamefont {Sokolovskii},\ and\
  \citenamefont {Kuchko}}]{vasiliev_spin_2007-1}%
  \BibitemOpen
  \bibfield  {author} {\bibinfo {author} {\bibfnamefont {S.~V.}\ \bibnamefont
  {Vasiliev}}, \bibinfo {author} {\bibfnamefont {V.~V.}\ \bibnamefont
  {Kruglyak}}, \bibinfo {author} {\bibfnamefont {M.~L.}\ \bibnamefont
  {Sokolovskii}}, \ and\ \bibinfo {author} {\bibfnamefont {A.~N.}\ \bibnamefont
  {Kuchko}},\ }\href {\doibase 10.1063/1.2740339} {\bibfield  {journal}
  {\bibinfo  {journal} {J. Appl. Phys.}\ }\textbf {\bibinfo {volume} {101}},\
  \bibinfo {pages} {113919} (\bibinfo {year} {2007})}\BibitemShut {NoStop}%
\bibitem [{\citenamefont {Gurevich}\ and\ \citenamefont
  {Melkov}(1996)}]{gurevich_magnetization_1996}%
  \BibitemOpen
  \bibfield  {author} {\bibinfo {author} {\bibfnamefont {A.~G.}\ \bibnamefont
  {Gurevich}}\ and\ \bibinfo {author} {\bibfnamefont {G.~A.}\ \bibnamefont
  {Melkov}},\ }\href@noop {} {\emph {\bibinfo {title} {Magnetization,
  {{Oscillations}} and {{Waves}}}}}\ (\bibinfo  {publisher} {{CRC Press Inc}},\
  \bibinfo {address} {New York},\ \bibinfo {year} {1996})\BibitemShut {NoStop}%
\bibitem [{\citenamefont {Landau}\ and\ \citenamefont
  {Lifshitz}(1960)}]{landau_electrodynamics_1960}%
  \BibitemOpen
  \bibfield  {author} {\bibinfo {author} {\bibfnamefont {L.~D.}\ \bibnamefont
  {Landau}}\ and\ \bibinfo {author} {\bibfnamefont {E.~M.}\ \bibnamefont
  {Lifshitz}},\ }\href@noop {} {\emph {\bibinfo {title} {Electrodynamics of
  {{Continuous Media}}}}}\ (\bibinfo  {publisher} {{Pergamon Press Ltd.}},\
  \bibinfo {address} {Oxford},\ \bibinfo {year} {1960})\BibitemShut {NoStop}%
\bibitem [{\citenamefont {Landau}\ and\ \citenamefont
  {Lifshitz}(1935)}]{landau_theory_1935-1}%
  \BibitemOpen
  \bibfield  {author} {\bibinfo {author} {\bibfnamefont {L.~D.}\ \bibnamefont
  {Landau}}\ and\ \bibinfo {author} {\bibfnamefont {E.}~\bibnamefont
  {Lifshitz}},\ }\href@noop {} {\bibfield  {journal} {\bibinfo  {journal}
  {Phys. Z. Sowjetunion}\ }\textbf {\bibinfo {volume} {8}},\ \bibinfo {pages}
  {101} (\bibinfo {year} {1935})}\BibitemShut {NoStop}%
\bibitem [{\citenamefont {Gorobets}\ \emph {et~al.}(1998)\citenamefont
  {Gorobets}, \citenamefont {Kuchko},\ and\ \citenamefont
  {Vasil'ev}}]{gorobets_excitation_1998}%
  \BibitemOpen
  \bibfield  {author} {\bibinfo {author} {\bibfnamefont {Y.~I.}\ \bibnamefont
  {Gorobets}}, \bibinfo {author} {\bibfnamefont {A.~N.}\ \bibnamefont
  {Kuchko}}, \ and\ \bibinfo {author} {\bibfnamefont {S.~V.}\ \bibnamefont
  {Vasil'ev}},\ }\href@noop {} {\bibfield  {journal} {\bibinfo  {journal}
  {Phys. Met. Metallogr.}\ }\textbf {\bibinfo {volume} {85}},\ \bibinfo {pages}
  {272} (\bibinfo {year} {1998})}\BibitemShut {NoStop}%
\bibitem [{\citenamefont {Wieser}\ \emph {et~al.}(2009)\citenamefont {Wieser},
  \citenamefont {Vedmedenko},\ and\ \citenamefont
  {Wiesendanger}}]{wieser_quantized_2009}%
  \BibitemOpen
  \bibfield  {author} {\bibinfo {author} {\bibfnamefont {R.}~\bibnamefont
  {Wieser}}, \bibinfo {author} {\bibfnamefont {E.~Y.}\ \bibnamefont
  {Vedmedenko}}, \ and\ \bibinfo {author} {\bibfnamefont {R.}~\bibnamefont
  {Wiesendanger}},\ }\href {\doibase 10.1103/PhysRevB.79.144412} {\bibfield
  {journal} {\bibinfo  {journal} {Phys. Rev. B}\ }\textbf {\bibinfo {volume}
  {79}},\ \bibinfo {pages} {144412} (\bibinfo {year} {2009})}\BibitemShut
  {NoStop}%
\bibitem [{\citenamefont {Gonz{\'a}lez}\ \emph {et~al.}(2010)\citenamefont
  {Gonz{\'a}lez}, \citenamefont {Landeros},\ and\ \citenamefont
  {N{\'u}{\~n}ez}}]{gonzalez_spin_2010}%
  \BibitemOpen
  \bibfield  {author} {\bibinfo {author} {\bibfnamefont {A.~L.}\ \bibnamefont
  {Gonz{\'a}lez}}, \bibinfo {author} {\bibfnamefont {P.}~\bibnamefont
  {Landeros}}, \ and\ \bibinfo {author} {\bibfnamefont {{\'A}.~S.}\
  \bibnamefont {N{\'u}{\~n}ez}},\ }\href {\doibase 10.1016/j.jmmm.2009.10.010}
  {\bibfield  {journal} {\bibinfo  {journal} {J. Magn. Magn. Mater.}\ }\textbf
  {\bibinfo {volume} {322}},\ \bibinfo {pages} {530} (\bibinfo {year}
  {2010})}\BibitemShut {NoStop}%
\bibitem [{\citenamefont {Borys}\ \emph {et~al.}(2016)\citenamefont {Borys},
  \citenamefont {Garcia-Sanchez}, \citenamefont {Kim},\ and\ \citenamefont
  {Stamps}}]{borys_spin-wave_2016}%
  \BibitemOpen
  \bibfield  {author} {\bibinfo {author} {\bibfnamefont {P.}~\bibnamefont
  {Borys}}, \bibinfo {author} {\bibfnamefont {F.}~\bibnamefont
  {Garcia-Sanchez}}, \bibinfo {author} {\bibfnamefont {J.-V.}\ \bibnamefont
  {Kim}}, \ and\ \bibinfo {author} {\bibfnamefont {R.~L.}\ \bibnamefont
  {Stamps}},\ }\href {\doibase 10.1002/aelm.201500202} {\bibfield  {journal}
  {\bibinfo  {journal} {Adv. Electron. Mater.}\ }\textbf {\bibinfo {volume}
  {2}},\ \bibinfo {pages} {1500202} (\bibinfo {year} {2016})}\BibitemShut
  {NoStop}%
\bibitem [{Note1()}]{Note1}%
  \BibitemOpen
  \bibinfo {note} {We use the following parameters throughout this paper: $M_0$
  = 800 erg$\cdot $G$^{-1}\cdot $cm$^{-3}$, $\gamma = 1.76\times 10^7$ Hz$\cdot
  $Oe$^{-1}$, $\alpha = 3.125 \times 10^{-12}$ cm$^{2}$, $\beta _\parallel =
  0.1$ ($K=32\times 10^3$erg$\cdot $cm$^{-3}$), $\beta _\perp = 10$
  ($K=32\times 10^5$erg$\cdot $cm$^{-3}$)}\BibitemShut {NoStop}%
\bibitem [{Note2()}]{Note2}%
  \BibitemOpen
  \bibinfo {note} {See Supplemental Material at [link TBC]}\BibitemShut
  {NoStop}%
\bibitem [{\citenamefont {Boone}\ and\ \citenamefont
  {Krivorotov}(2010)}]{boone_magnetic_2010}%
  \BibitemOpen
  \bibfield  {author} {\bibinfo {author} {\bibfnamefont {C.~T.}\ \bibnamefont
  {Boone}}\ and\ \bibinfo {author} {\bibfnamefont {I.~N.}\ \bibnamefont
  {Krivorotov}},\ }\href {\doibase 10.1103/PhysRevLett.104.167205} {\bibfield
  {journal} {\bibinfo  {journal} {Phys. Rev. Lett.}\ }\textbf {\bibinfo
  {volume} {104}},\ \bibinfo {pages} {167205} (\bibinfo {year}
  {2010})}\BibitemShut {NoStop}%
\bibitem [{\citenamefont {Schl{\"o}mann}(1964)}]{schlomann_generation_1964-1}%
  \BibitemOpen
  \bibfield  {author} {\bibinfo {author} {\bibfnamefont {E.}~\bibnamefont
  {Schl{\"o}mann}},\ }\href {\doibase 10.1063/1.1713058} {\bibfield  {journal}
  {\bibinfo  {journal} {J. Appl. Phys.}\ }\textbf {\bibinfo {volume} {35}},\
  \bibinfo {pages} {159} (\bibinfo {year} {1964})}\BibitemShut {NoStop}%
\bibitem [{\citenamefont {Li}\ \emph {et~al.}(2002)\citenamefont {Li},
  \citenamefont {Lew}, \citenamefont {Bland}, \citenamefont {Lopez-Diaz},
  \citenamefont {Natali}, \citenamefont {Vaz},\ and\ \citenamefont
  {Chen}}]{li_microstructures:_2002}%
  \BibitemOpen
  \bibfield  {author} {\bibinfo {author} {\bibfnamefont {S.~P.}\ \bibnamefont
  {Li}}, \bibinfo {author} {\bibfnamefont {W.~S.}\ \bibnamefont {Lew}},
  \bibinfo {author} {\bibfnamefont {J.~A.~C.}\ \bibnamefont {Bland}}, \bibinfo
  {author} {\bibfnamefont {L.}~\bibnamefont {Lopez-Diaz}}, \bibinfo {author}
  {\bibfnamefont {M.}~\bibnamefont {Natali}}, \bibinfo {author} {\bibfnamefont
  {C.~A.~F.}\ \bibnamefont {Vaz}}, \ and\ \bibinfo {author} {\bibfnamefont
  {Y.}~\bibnamefont {Chen}},\ }\href {\doibase 10.1038/415600a} {\bibfield
  {journal} {\bibinfo  {journal} {Nature}\ }\textbf {\bibinfo {volume} {415}},\
  \bibinfo {pages} {600} (\bibinfo {year} {2002})}\BibitemShut {NoStop}%
\end{thebibliography}

\end{document}